\documentclass{LT23auth}
\usepackage{graphicx}

\begin{document}

\begin{frontmatter}

\title{The Anderson transition
due to random spin-orbit coupling in two-dimension}

\author[address1]{Yoichi Asada \thanksref{thank1}},
\author[address1]{Keith Slevin},
\author[address2]{Tomi Ohtsuki}

\address[address1]{Department of Physics, Graduate School of Science,
Osaka University, 1-1 Machikaneyama, Toyonaka, Osaka 560-0043, Japan}

\address[address2]{Department of Physics,
Sophia University, Kioi-cho 7-1, Chiyoda-ku, Tokyo 102-8554, Japan}

\thanks[thank1]{ E-mail:asada@presto.phys.sci.osaka-u.ac.jp}

\begin{abstract}
We report an analysis of the Anderson transition in an SU(2) model with chiral
symmetry. 
Clear single parameter scaling behaviour is observed.
We estimate the critical exponent for the divergence
of the localization length to be $\nu=2.72\pm.02$
indicating that
the transition belongs to the symplectic universality class.
\end{abstract}
%
%
\end{frontmatter}

\section{Introduction}
It is thought that the critical phenomena of the Anderson transition
may be classified according to three universality classes
(orthogonal, unitary and symplectic)
depending on the symmetries of the Hamiltonian,
i.e. time reversal and spin rotation symmetries.
A different value of the critical exponent
$\nu$ describing the divergence of the localization length
is expected to characterize each universality class.

In the absence of any diagonal disorder
the SU(2) model Hamiltonian may have chiral symmetry.
This depends on the boundary conditions
and the number of sites in each direction.
Near the band center chiral symmetry affects the
localization of electrons,
e.g., an even-odd system size dependence of the quasi-1d
localization length is observed \cite{furusaki}.
Here we investigate whether or not
chiral symmetry also affects the critical exponent $\nu$.

A system belongs to the symplectic 
universality class if its Hamiltonian commutes with
a time reversal operator $\mathcal{T}$ that satisfies
$\mathcal{ T}^2=-1$, e.g., systems with significant spin-orbit coupling.
Systems in this universality class  
exhibit an Anderson transition even in 2D.
Recently we analysed the SU(2) model with an on-site random
potential (diagonal disorder), which breaks chiral symmetry,
and estimated the critical exponent $\nu=2.73\pm.02$ \cite{asada}.

Here we study the SU(2) model without an on-site random potential.
In spite of the fact that there is no random potential, we find
that there is an Anderson transition at a critical energy $E_c$.
Further this energy $E_c$ is far from the the band center
and chiral symmetry does not affect the observed critical phenomena.

\section{Model and Method}
The Hamiltonian of the SU(2) model describes non-interacting electrons
on a simple square lattice with nearest neighbour
SU(2) random hopping
\begin{eqnarray}
H
=
-
\!\!\!\!\!
\sum_{\langle i,j \rangle,\sigma,\sigma'}
\! \!
\left(
\! \!
\begin{array}{cc}
e^{i\alpha_{ij}}\! \cos \beta_{ij}
& e^{i\gamma_{ij}}\! \sin \beta_{ij} \\
-e^{-i\gamma_{ij}}\! \sin \beta_{ij}
& e^{-i\alpha_{ij}}\! \cos \beta_{ij}
\end{array}
\! \!
\right)
_{\!\!\! \sigma \sigma'}
\!\!\!\!\!
c_{i\sigma}^{\dagger}c_{j\sigma'}
\label{hamiltonian}
\end{eqnarray}
where $c^{\dagger}_{i\sigma}$($c_{i\sigma}$)
denotes the creation (annihilation)
operator of an electron at the site $i$ with spin $\sigma$.
We distribute hopping matrices randomly and independently
with uniform probability on the group SU(2).
This corresponds to $\alpha$ and $\gamma$ uniformly distributed
in the range $[0,2\pi)$,
and $\beta$ distributed in the range $[0,\pi/2]$
according to the probability density,
$P(\beta){\rm d}\beta=
\sin (2\beta) {\rm d}\beta$.

We calculate the localization length $\lambda$ on a quasi-1d strip
whose width is $L$ with the transfer matrix method
\cite{mackinnon} and analyse
the dependence of the re-normalized localization length
$\Lambda = \lambda/L$
on the energy and the width of the strip
\cite{slevin}.
We also calculate the density of states
with the Green's function iteration method \cite{schweitzer}.

In the course of our calculations we noticed a technical
problem with the method described in \cite{mackinnon}.
When fitting the numerical data it is important to have
not only an estimate of the localization length $\lambda$
but also an estimate $\delta \lambda$ of how accurately
$\lambda$ has been estimated.
Otherwise a reliable fitting of the data is not possible.
We have found that the error $\delta \lambda$ is over
estimated by the method of \cite{mackinnon}.
To overcome this problem we divide the quasi-1d strip
into blocks whose lengths are $10^3$
and calculate the quasi-1d localization length
$\lambda_i$($i$ denotes the block) in each block
and accumulate $\lambda_i$.
From the distribution of $\{\lambda_i\}$
we estimate $\lambda$ and $\delta \lambda$.


\begin{figure}[tbp]
\begin{center}\leavevmode
\includegraphics[width=0.8\linewidth]{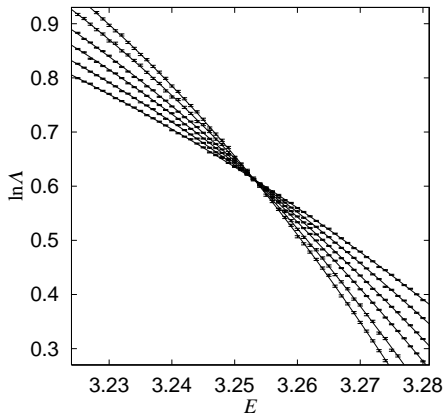}
\caption{
$\ln \Lambda$ vs Fermi energy $E$
for system sizes $L=11,16,23,32,47,64$.
The solid lines are the fit.
}
\label{fig:su2w0cp}\end{center}
\end{figure}

\section{Results}

We calculated $\Lambda$ as a function of Fermi energy
for systems with widths $L=11, 16, 23, 32, 47$ and $64$.
The  accuracy of the localization length data is $0.1\%$,
with the exceptions of $L=47,64$ where the accuracy is $0.2\%$.
To achieve this accuracy
systems of length of the order of $10^7$ to $10^8$ are required.
We impose the periodic boundary conditions in the 
transverse direction.
For this choice of boundary condition systems with
even $L$ have chiral symmetry, while those with odd $L$
do not.
We fit even and odd data separately first and then
in combination.
The results are displayed in Table \ref{table:result}.
The estimates of the critical parameters for all three
fits are in close coincidence and we conclude that
chiral symmetry does not change the universality class.
In Fig. \ref{fig:su2w0cp} we show the result for the
combined even and odd data.
The lines cross at single point,
indicating both that corrections to scaling
due to irrelevant variables
are negligible and also that an even-odd effect does not exist.
The estimates $\nu$ and $\ln \Lambda_c$
are consistent with those in \cite{asada}
for the symplectic universality class.

\begin{figure}[tbp]
\begin{center}\leavevmode
\includegraphics[width=0.8\linewidth]{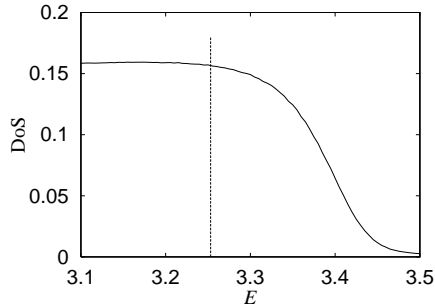}
\caption{
The DoS near the band edge
for a system of size $100\times 100$.
The vertical line shows the mobility edge ($E_c\simeq 3.2531$).
(The imaginary part of the energy is 0.005.)
}
\label{fig:su2w0dos}\end{center}
\end{figure}

The density of states
for a system of the size $100\times 100$
averaged over $10^3$ ensembles is shown in
Fig.\ref{fig:su2w0dos}.
In the critical region, the density of states changes
slowly.
This behaviour is different from that of the random magnetic
flux U(1) model in 3D where the density of states changes rapidly
in the critical region \cite{ohtsuki}.
This might be the reason why we obtain clearer
single parameter scaling here than in a study of the
critical phenomena of the U(1) model \cite{kawarabayashi}.


\begin{table}
\caption{
The number of data $N_d$ and goodness of fit $Q$.
The best fit estimates of the critical energy $E_c$,
$\ln\Lambda_c$ and the critical exponent $\nu$
with $95\%$ confidence intervals.
There are 7 fitting parameters.
}
\label{table:result}
\begin{tabular}{p{8mm}p{7mm}p{6mm}p{20mm}p{17mm}p{13mm}}
\hline
    &$N_d$& $Q$ & $E_c$            & $\ln\Lambda_c$   & $\nu$        \\
\hline
Even& 159 & 0.1 & 3.2531$\pm$.0002 & 0.613$\pm$.002   & 2.72$\pm$.02 \\
Odd & 164 & 0.2 & 3.2531$\pm$.0002 & 0.613$\pm$.002   & 2.72$\pm$.03 \\
Both& 323 & 0.1 & 3.2531$\pm$.0001 & 0.613$\pm$.001   & 2.72$\pm$.02 \\
\hline
\end{tabular}
\end{table}


%
%
%

%
%

\end{document}